# Research Progress of Terahertz Technology in Microbiology

Ding Cao[1,2], Guangyou Fang[1,2], Xuequan Chen[1,2,*]

[1] GBA Branch of Aerospace Information Research Institute, Chinese Academy of Sciences, Guangzhou, 510700, China
[2] Guangdong Provincial Key Laboratory of Terahertz Quantum Electromagnetics, Guangzhou, 510700, China
* corresponding author: chenxq@aircas.ac.cn

**Abstract**

Microorganisms are ubiquitous in nature, and microbial activities are closely intertwined with the entire life cycle system and human life. Developing novel technologies for the detection, characterization and manipulation of microorganisms promotes their applications in clinical, environmental and industrial areas. Over the last two decades, terahertz (THz) technology has emerged as a new optical tool for microbiology. The great potential originates from the unique advantages of THz waves including the high sensitivity to water and inter-/intra-molecular motions, the non-invasive and label-free detecting scheme, and their low photon energy. THz waves have been utilized as a stimulus to alter microbial functions, or as a sensing approach for quantitative measurement and qualitative differentiation. This review specifically focuses on recent research progress of THz technology applied in the field of microbiology, including two major parts of THz biological effects and the microbial detection applications. In the end of this paper, we summarize the research progress and discuss the challenges currently faced by THz technology in microbiology, along with potential solutions. We also provide a perspective on future development directions. This review aims to build a bridge between THz photonics and microbiology, promoting both fundamental research and application development in this interdisciplinary field.

# 1. Background and Introduction

The earth is flooded with micron-sized microorganisms. Although microorganisms cannot be seen by naked eyes, they interact with human beings at every moment. For example, an adult lives with ~$10^{24}$ microorganism together in gut, mouth and skin [1, 2], of which amount is even larger than human cells [3]. Microbes are not only able to cause diseases, but also alter human appetite and regulate food intake by releasing proteins [4] that activate anorexic neurons in brain. Detecting microorganisms, characterizing their cellular components and investigating their microbial behaviors under various stimuli are vital for agricultural, clinical and industrial settings.

Microorganisms contain all micro-size living matters from the three-domains system (i.e., *Archaea*, *Bacteria* and *Eukarya*). *Archaea* and *Bacteria* are collectively termed as prokaryotes, defined as single-cell organism that lacks of nucleus and other membrane-bound organelles. Prokaryotes are considered as more primitive than eukaryotes. *Eukarya* has membrane-bound organelles and includes animals, plants, fungi and other membrane-bound protist. A portion of *Eukarya* live in unicellular form and are consider as microorganisms. Both unicellular form (microns in length) and colony form (larger than centimeter) are common morphologies of microorganisms in daily life. In addition, although viruses are not generally considered as living matters and are not classified into the three-domain system, THz virus research adopts similar protocols to those used for living microorganisms considering their comparable size. Therefore, we will consider viruses as a part of microorganisms in this review.

Although there have been numerous technologies for studying microorganisms, there are still many yet-unknown properties and application potential, highlighting the continued need for novel research tools. In recent decades, a less-explored electromagnetic spectrum, the THz regime ($10^{11}$ Hz to $10^{13}$ Hz), has become an emerging research area. Scientists are curious about how THz waves interact with matters (especially bio-materials), since THz radiation is missing from natural sunlight due to the strong atmosphere absorption (mainly water vapor). The interactions of microorganisms with other spectral radiations are better understood. For example, infrared light can enhance cell proliferation [5] whilst ultraviolet light can destroy cellular structures [6]. Studies on the interaction between THz radiation and microorganisms may reveal some unseen biological effects during cellular growth.

THz waves locate in between microwaves and infrared, as shown in Fig.1. Throughout the electromagnetic spectrum, each frequency range has found their unique applications, such as wireless communications by microwaves and radio waves, visual applications by visible light and medical diagnosis by X-ray. In contrast, THz waves are less studied due to the historical lack of efficient sources and sensitive detectors. With advancing instrumentations, THz photonic technology has become more widespread nowadays, especially in biomedicine. THz waves have similar energy levels with intermolecular motions of biomolecules, making THz spectroscopy sensitive to characteristic phonon absorption of biomaterials. THz frequencies also correspond to hydrogen-bonding relaxations; hence water presents strong dispersion and absorption in the THz range, highly contrasting that of macro-biomolecules.

Although there have been a few THz biomedical reviews focused on mammalian cells and tissues microorganisms have rarely been comprehensively compared, summarized, and discussed. This review aims to fill the gap in this area by conducting a systematic investigation, organization, analysis, and outlook on THz-related research in microbiology. Our goal is to provide valuable insights for researchers in related areas of THz photonics, microbial biology, and biophysics. This paper is organized as follows. In Chapter 2, we will summarize findings about the biological effects of microorganisms under THz radiation, taking THz waves as stimulus to investigate how microorganisms response to the radiation. Chapter 3 will discuss detecting technologies developed for microorganisms based on THz spectroscopy. Finally, we discuss the current challenges and envision future outlooks in the last chapter.

## 2. Biological Effects of Microorganisms under THz Radiation

THz technology has seen rapid progress in biomedical applications. The development of high-power THz sources is one of the driving forces. Nowadays, intense THz radiation can be generated through various approaches, including solid-state devices, spintronic devices [7], non-linear crystals [8] and free electron lasers [9]. Even fiber-coupled photoconductive antennas can reach milliwatt-level output power [10]. The effects of THz radiation on humans and biological organisms have attracted increasing scientific interest for several reasons. First, THz radiation has rarely been involved in the biological evolution process because of the strong absorption by the atmosphere. How creatures response to THz waves remains an open question. It is known that the high-water concentration of tissues significantly attenuates the THz light to a penetration depth of only a few hundreds of micrometers, hence the radiation influence on animals is limited to the upper skin layers. However, such thickness is already beyond the size of microorganism and could completely change its behavior. Second, curiosity stems from the unique properties of the THz wave. Specifically, THz waves are absorbed differently by different biological molecules, such as DNA and proteins. The frequency of THz waves coincides with the long-range collective intermolecular motions of micro-biomolecules and the intermolecular vibrations of macromolecules. The energy level of these waves corresponds to the energy of hydrogen bonds and Van der Waals force, as illustrated in Fig 1. Owing to these interactions, THz waves can induce both thermal and non-thermal biological effects, positioning them as promising tools for modifying the configuration of biological molecules and the physiological state of living cells.

In 2015, the Scientific Committee on Emerging and Newly Identified Health Risks (SCENIHR), guided by the European Commission, shared its detailed opinion on the 'Potential health effects of exposure to electromagnetic fields' [11]. This important document highlights that as THz technology advanced, there was a growing need to study the potential biological effects of THz exposure. SCENIHR specifically recommended more research on the effects of THz radiation on the skin (for long-term and low-level exposure) and on the cornea (for short-term and high-intensity exposure). Over the past decades, researchers have investigated the effects of THz radiation on skin cells, nerve cells, corneal tissues and so on. For example, some studies found increased gene expression and cell proliferation in skin cells after THz exposure [12-15], whilst some studies suggest no radiation effect [16-18]. Changes in gene expression [19], cell membrane properties [20], and overall cell viability [21] have also been reported on nerve cells. A study by Zhao et al. [19] exposed mice cortical neurons to THz radiation and found increased neurite outgrowth and synaptic-associated gene expressions, leading to increased excitatory synaptic transmission and neuronal firing.

In comparison to mammalian cells, microorganisms typically have higher cell count [1, 2], exhibit greater cellular diversity [22], and possess a simpler cellular structure [23]. Their existence can manifest as single-celled entities or as complex colony communities, with their form often influenced by species type, growth phase and environmental conditions [24-26]. Notably, a single-celled microorganism contains a complete set of genetic materials, enabling it to survive and reproduce independently. In contrast, mammalian cells operate in a more interconnected manner, described metaphorically as 'cogs in a machine', working in unison to create functional organs. Many microorganisms are easy to culture, and exhibit strong viability and practicability in agricultural [27], industrial [28] and clinical settings [29]. Furthermore, many studies about biological effects on mammalian organisms are restricted to the outermost layer (hundreds of micrometers) due to the limited penetration depth of THz waves in highly hydrated layers. This limitation makes it challenging to study the comprehensive effects of THz radiation on entire tissues, organs, or living organisms. In contrast, microorganisms can be fully penetrated by THz waves, leading to altered gene expression and modified cellular behaviors. Despite their significance, our understanding of the interactions between microorganisms and THz radiation remains limited, especially when compared to the knowledge amassed on mammalian cells. This research gap highlights the need for more targeted research in this area. In the following of this chapter, we will review existing literature on the biological effects of THz radiation on microorganisms, compare these findings with those on mammalian cells and discuss the potential future directions this research might take. In order to maintain the fluency of the article and facilitate better cross-comparison, the diverse experimental conditions including THz sources and types, radiation frequencies and intensity, as well as exposure time and experimental temperature are summarized in Table 1 rather than elaborated upon in the main text.

2.1. Direct Biological Effects on Microorganisms

Exposure to THz radiation has been observed to stimulate gene expression and enhance cell metabolism in mammalian cells [13, 15, 21, 30]. For microorganisms, bacteria *Escherichia coli* (commonly abbreviated as *E. coli*; a model organism in molecular biology) has been mostly studied to understand the biological effects of THz radiation. Numerous findings indicated that THz radiation can affect bacterial metabolism. For example, in a series of experiments by Peltek et al. [31] focusing on the impact of THz radiation on gene activities in *E. coli*, they revealed an upregulation in genes associated with cell aggregation and adhesion (Fig. 2a-b), while those related to cell division showed decreased activity. In another study [32], *E.coli* were exposed to a 3.1 THz continuous wave for an 8-hour duration, and an increment of the plasmids copy number was observed, which subsequently led to a rise in the production of red fluorescence protein (Fig. 2c-e). Shifting the focus to the transcriptional level, Imashimizu et al. [33] centered their research on the RNA polymerase activity of *E.coli* under THz radiation. Their results indicated significant impact in transcription processes, including abortive initiation and pausing. Focusing on protein translation, Ivanova et al. [34] found bacteria *E. coli* enable cellular responses to osmotic stress, plasma membrane regulation and phospholipid biosynthetic process under THz radiation from a broadband synchrotron source of 0.5−18 THz. After THz radiation, deformed outer membrane, membrane perturbations and leakage of cytosol of bacteria were observed (Fig. 2f-g). Studies have also been conducted on bacteria other thant *E. coil*. Bannikova et al. [35] explored the non-thermal THz effects on the extremophilic bacterium *Geobacillus icigianus*. They observed changes in various metabolic pathways, such as chemotaxis and the synthesis of peptidoglycan and riboflavin, after both short-term and long-term THz exposures. Recently, they found that the biological effects on *G. icigianus* were mainly attributed to disturbances in the expression of genes of the copper, iron and zinc homeostatic systems [36]. One advantage of working with thermophilic strains (such as *G. icigianus*) is their inherent resistance to temperature changes, allowing researchers to better isolate the radiation effects from thermal influence. Otherwise, additional temperature monitoring or control is necessary, following procedures described in ref [31-34].

Although the above studies show that THz waves can alter bacterial cellular activities and the inherent mechanisms at the molecular level are still unclear, there is no clear evidence that THz waves cause genetic mutations inside bacterial cells. For instance, Sergeeva et al. [37] explored the potential mutagenic and genotoxic effects of 2.3 THz radiation on bacteria, concluding that THz waves do not pose harm to bacterial cells. In a more comprehensive study, Shirato et al. [38] employed the Ames test (a widely employed method of evaluating the potential of a stimuli to cause DNA mutations of bacteria) on five different bacterial strains, subjecting them to various stimuli, including 1.6 THz pulse laser, UV radiation, and chemical stimulants. Their findings consistently indicated that THz radiation did not exhibit mutagenic properties, nor did it inflict DNA damage (Fig. 3a-b). In contrast, obvious DNA damage (Fig. 3a) and low cell viability (Fig. 3b) were observed after UV radiation and chemical stimulants. Therefore, the photon energy of THz waves is confirmed to be insufficient to break molecular bonds and cause genetic mutations. THz waves are more likely to interact with the gene duplication, transcription or translation process to up/down regulate some biomolecular activities.

While most researchers have focused on bacteria, studies on the effects of other species have also been reported. Hadjiloucas et al. [39] conducted preliminary tests on yeast cells, a type of fungi, exposing them to radiation frequencies between 200-350 GHz. They observed an increased growth rate specifically at 341 GHz. In another study, Goryachkovskaya et al. [40] found that the expression levels of 16 proteins in Archaea were altered when exposed to THz radiation. Additionally, diatom algae, a type of phytoplankton with a protective siliceous layer (known as a frustule), showed a unique response that the separation of frustules from diatom's cell membrane can be promoted by submillimeter-wave radiation (Fig. 3c-e) [41].

2.2. Utilization of Biological Effects
In addition to studying the biological effects induced by THz radiation, scientists have utilized bacteria as THz-sensitive biosensors to examine metabolic pathways within bacterial cells. Note that the biosensors (short for 'biological sensors') have a different definition to THz optical sensors for sensing biomedical analytes. Here, the biosensors refer to bacteria with proper transducers that response to targeted substances by producing easily measured physicochemical signals under THz stimulus. Many of these biosensors are based on the bacteria *E.coli* [42-45], due to its well-known genetic information. For example, through genetic engineering, genes that are sensitive to heavy metals can be linked with genes that express fluorescence in a plasmid. This allows bacteria-based biosensors to detect heavy metals by observing bacterial fluorescence. Since 2013, Peltek et al. have been working on building *E. coli* biosensors to interact with THz waves. Their approach involves using a plasmid containing a promoter and a reporter gene. The promoter is a DNA sequence that initiates DNA transcription, and the downstream reporter gene is a DNA sequence to be transcribed (e.g. the green fluorescent protein GFP). THz waves can interact with the promoter and increase the expression level of the reporter gene. This setup allows the observation of specific gene expressions in real-time and *in situ* when exposed to THz radiation. For example, promoter *katG* gene encodes for hydroperoxidase I, which protects aerobic and phosphate-starved cells from oxdative damage [46, 47]. *pKatG-GFP* can thus act as a specific indicator for oxidative damage to *E. coli*. Demidova et al. [48] transformed plasmid *pKatG-GFP* into bacteria *E.coli* and found bacteria expressed more GFP with THz radiation (Fig. 4a-c). The increased GFP indicated higher expression of hydroperoxidase I, which is the primary catalase in the hydrogen peroxide-degrading metabolic pathways in bacterial cells [49]. This *E. coli* biosensor can be used to study the metabolism dynamics of *katG* related genes. Furthermore, the promoter of other sensor genes (including *copA*, *emrR*, *matA*, *safA*, *chbB* and *tdcR*) have been used to investigate different metabolic pathways [49-52], showing similar increasing gene expressions under THz radiation (Fig. 4d-e).

2.3. Summary

THz radiation has demonstrated many unique non-thermal effects on microorganisms, showing different characteristics compared to other stimulus. However, this area is still in its early stage and lacks comprehensive and systematic research. We summarize the literature mentioned above in Table 1, including their experimental conditions. The table shows obviously different controlling parameters (power, frequency, illumination time etc.) and investigated biological effects, making it challenging to make a comparison or definitive conclusions. Some preliminary findings can be observed. For example, the response of microorganisms to THz radiation depends on the radiation parameters, including the center frequency, exposure time and power density. Given that THz frequencies align with the collective vibrational modes of many biomolecules, and their energy levels match those of hydrogen bonds and Van der Waals interactions, it is reasonable to expect that THz waves interact with molecules within microorganism cells and have subsequent impact on microbial life. Therefore, additional experiments should be done to reveal the fundamental response and mechanisms.

Most literature shows positive cellular response (increasing DNA and protein expression) with respect to THz radiation. Future studies may utilize this characteristic to activate microbial cellular functions to increase the rate of reproduction or microbial products. In fact, the majority of discovered microorganisms are nonculturable or difficult to cultivate in laboratory environments [53]. This type of microorganism is known as viable but nonculturable cells (VBNCs). VBNCs are normally in a low metabolic state without growth and proliferation, but they become active and culturable once resuscitated. How to trigger cell resuscitation is critical for obtaining more resources of culturable microbial strains and their genetic information. Traditional research of VBNC has been focused on modifying bacterial cultivation conditions by imitating the natural environment and adding small molecules that may be essential for bacterial growth. Here, THz waves demonstrate their potential importance in culturing VBNC microorganisms, serving as a non-destructive tool to interact with specific gene sequence or proteins to initiate resuscitation.

It should also be noted that some studies suggested that THz radiation does not affect bacterial viability [54] or their metabolic activities [55], while exposure to approximately 6 J of total THz wave energy could even damage bacterial cell and cause cell death [56]. These effects should also be studied and considered in practical applications. While groundbreaking discoveries in this area might still be on the horizon, the potential of THz technology in microbiology remains promising and could reveal unforeseen insights and applications.

## 3. Detection of Microorganisms Using THz Waves

Although many accurate and target-selective detection methods for biomedical applications are readily available, such as quantitative polymerase chain reaction (qPCR), they are generally time-consuming, non in situ, destructive, and often requiring experimental expertise. Spectroscopy is a promising tool for analyzing chemical and biological matters non-invasively and rapidly, which commonly uses frequencies spanning from ultraviolet to infrared. THz waves lie at the far-infrared range that provide unique information of long-range interaction, including lattice phonon vibrations of nucleotide, amino acids and carbohydrates, as well as intramolecular vibrations of large molecules such as proteins. The rich hydrogen-bond network also responses to THz waves, showing characteristic high absorption and dispersion. THz spectroscopy offers fast, label-free, non-destructive and quantitative detection ability, exhibiting promising potentials in material science, food industry, agriculture and biomedicine. In this chapter, we will focus on the realm of microorganism and review their detection using THz technologies.

3.1. Instrumentations and sensors

3.1.1. Experimental setup

THz systems can be classified into pulsed time-domain THz systems and continuous-wave (CW) THz systems. The former is often known as THz time-domain spectroscopy (THz-TDS) systems, which generate and detect THz radiation via photoconductive antennas, non-linear crystals or spintronic devices excited by femtosecond infrared light. The picosecond-scale single-cycle THz electric field is detected in time-domain. THz-TDS systems are featured by their picosecond time-resolution, ultrabroad bandwidth and coherent detection that provides both magnitude and phase information simultaneously. CW THz systems have various source-detector combinations. They usually have narrow linewidth and high output power, hence providing better spectral resolution and deeper penetration depth. A comprehensive overview of THz instrumentation is beyond the scope of this review, but can be found from other literature [57].

Both pulsed and CW THz systems can be built in two fundamental configurations of transmission or reflection, and both configurations have been used in microorganism detection. In transmission, THz beams transmitting through the holder or the device are measured with and without the investigated sample. The two signals are compared and analyzed to extract the intrinsic sample properties. Biological samples are usually hydrated that their high water-concentration strongly absorbs THz light and limits the use of transmission configuration to thick samples. In this case, reflection configuration can be used as it is not affected by the absorption. Samples are usually placed on a flat supporting medium to establish a well-focused interface. Amplitude and phase change of the reflected signal records the sample information.

### 3.1.2. Sensors

One of the biggest challenges of sensing microorganisms using THz waves is the low light-sample interaction efficiency due to the scale mismatch between the THz wavelengths and the sample size. The micrometer-size of microorganisms is over an order smaller than the sub-millimeter wavelengths of THz waves, resulting in weak changes to the THz signal that can be easily affected by noise and measurement errors. Therefore, sensors capable of enhancing the sample-interaction efficiency are highly demand. This is typically achieved by field-confining techniques. Metamaterials (MMs) are the most widely used type, which contain artificial sub-wavelength periodic structures resonating at specific frequencies. The resonant waves are strongly confined to the surface of the sensor, typically with a height of only a few micrometers. Such a design not only perfectly matches the scale of microorganisms but also significantly enhances the interaction efficiency and detection sensitivity. In the following review we found that about 40 of the total 60 reported works used MMs. Trace amounts of analytes can sensitively affect the resonant frequency, of which can be unambiguously detected. THz MMs have been extensively studied over the past two decades, and they play growing importance in microorganisms (or their components) detection. Apart from MMs, other strategies have also been reported, such as utilizing the evanescent mode of a suspended core THz fiber [58], or by using a metallic mesh sensor [59], antenna [60], or THz surface plasmon polaritons (SPPs) [61, 62].

We summarize MMs-based sensors for microorganism detection in Table. 2 due to their widespread application. The table clearly shows that the most widely used MM structure is the metallic split-ring resonator (SRR) structure, whilst other designs have also been reported. A common architecture is fabricating the metallic MM structure on THz-transparent substrates of silicon, quartz or polymer. Some additional techniques may be applied to further enhance the sensitivity, such as in combination with rolling circle amplification [63], nanoparticles [64], graphene and nano-antenna structures. For sample preparation, most experiments simply deposit and dry the analyte onto the sensor surface, whilst some studies functionalized the sensor with specific antibody to selectively detect targeted microorganisms, especially viruses. The resonance with specific THz frequencies is reflected as resonant peaks and dips in the spectrum, depending on the sensor types and optical configurations, both can be used to evaluate the sensor response. The red shift of the resonant frequency and the decay of the resonant strength are the typical responses of MMs to the loaded microorganisms, due to the increased dielectric constant and the damping effect caused by the increased absorption. As such, quality-factor (Q-factor) and sensitivity are the two most important parameters to evaluate the sensor performance. The former is defined as the ratio of the resonant frequency over the resonance width (typically full width at half maximum, FWHM). It describes how narrow the resonance is, with a sharper peak or dip enabling a more accurate readout of the resonant frequency. The sensitivity is usually defined as the frequency shift per unit refractive index change (RIU), or sometimes per unit volume/concentration. Note that these parameters should be evaluated together to estimate the general performance, and one should notice that there are usually big differences between the simulation results and the practical values. For example, a Q-factor up to a few hundred or even higher can be achieved in an ideal simulated structure, while practically Q-factor beyond 100 are barely reported due to the limitation in fabrication accuracy. Some studies have reported sensitivity over 1 THz/RIU at a very high frequency according to the simulation, which could be challenging to achieve since most THz systems cannot provide sufficient dynamic range at these frequencies.

3.1.3. Sample forms and preparation

Microorganisms naturally exist in single-cell form and colony form. Microbial single cell floats in liquid medium or adheres to surfaces without contacting other cells, while in the colony form cells are densely packed and fulfilled with extracellular polymeric substances (EPSs) [65]. Microorganisms growth often requires a humid or aqueous environment. Unfortunately, water is one of the most absorbing materials in the THz regime hence samples are usually dehydrated or prepared as thin films for THz measurements [57]. For example, single-cell bacteria are directly deposited on MM sensor surface and left dry. Bacterial colony is scraped from agar surface and filled into a defined-height chamber for THz measurements. Heating dehydration may be applied to further reduce the water content. The drying process may destroy the original structure of microorganism, making it no longer in situ and non-destructive. Freeze-drying is another frequently used method to reduce the absorption influence since ice is much less absorptive to THz waves. This approach better maintains the original cellular structures. Some microbial samples are naturally in a form appropriate for THz measurements. For example, leaf mildew is a disease of plant leaf infected by microorganisms [66]. The thin leaf (hundreds of microns) allows the penetration of THz waves, enabling distinguishing infected and non-infected areas based on their water content difference. In addition, with proper optical configurations, some samples were measured in highly hydrated states, which will be discussed in chapter 4.1.

3.2. Living microorganisms

3.2.1. THz spectroscopy of microorganisms

Spectroscopy of microorganisms has been widely studied in the visible and infrared region, including Fourier-transform infrared (FTIR) spectroscopy, fluorescent spectroscopy and Raman spectroscopy. As the extension of infrared, the photon energy of the THz waves matches the energy levels of low-frequency bio-molecular vibrations originating from intra- and inter-molecular weak interactions. It is expected that these molecular motions may specifically response to some of the THz frequencies to produce characteristic absorption features. However, they cannot be observed in targets with complex molecular compositions such as microorganisms, which has been explicitly discussed in the perspective by Markelz and Mittleman [67]. In detail, the intermolecular oscillations of small molecules can only produce observable features in the crystalline form from their phonon resonance, which are the result of the long-range ordering structure. Macro-molecules have a high density of states which result in featureless continua. Aqueous samples are featured by the continuous and strong absorption from highly damped hydrogen-bonding network. Studies on amorphous small molecules [68, 69], molecular solutions [70-72] and macro-biomolecules [73, 74] have validated the above physical predictions. For samples like microorganisms which contain a huge number of different molecules and often in aqueous state, there remains no possibility to observe any characteristic feature.

Indeed, Johnson et al. [75] measured the mid-IR and far-IR spectra of five *Bacillus* strains. They found characteristic frequencies in mid-IR range, but no signature in THz range (far-IR). Tang et al. [76] investigated the spectral response from 0.2 to 2.2 THz of *Bacillus* spores. Similarly, no THz signatures were observed for either bacterial cells or their main chemical components. The authors explained that the monotonic increased absorption with THz frequency was mainly contributed from Mie scattering and remnant water. Several studies also measured the absorption spectra of different bacteria [77, 78]. Although the slope of absorption-frequency curve differs in different bacterial samples, there were no characteristic frequency peaks that can be used as identification labels. Moreover, in studies of halophilic archaea, there is no signature in absorption-frequency curve of its main component bacteriorhodopsin [79, 80].

It should also be mentioned that some studies claim to have observed characteristic absorption of microorganisms in the THz band, including *E. coli* cells [81-83] and *Bacillus* spores [82-86]. Note that these reports are only from two groups. The characteristic frequencies of *E.coli* were found to be located at 14.5-14.9 $cm^{-1}$ and 17.1 $cm^{-1}$. For *Bacillus* species, signatures at different frequencies were reported, such as 250 GHz, 415 GHz and 1035 GHz for *Bacillus subtilis*, 955 GHz and 1015 GHz for *Bacillus thuringiensis*, 811 GHz, 830 GHz and 880 GHz for *Bacillus cereus*. These literatures suggested that the genetic materials [81, 83, 87] or the cellular major constitute (e.g., dipicolinic acid) [88] could be the source of the signatures. However, these explanations contradict with the condense matter theory. It is worth noting that all the above works used frequency-domain THz systems, which exhibit dense and complicated oscillations in the spectrum originated from the standing-wave effect, which can significantly interfere with the identification of characteristic absorptions. The observed peaks could also be induced by noise, errors or cavity-like oscillations, which should be carefully removed or calibrated to extract the intrinsic sample information, as highlighted in ref [67].

3.2.2. Bacterial detection and identification using THz waves

Instead of relying on the spectral signatures, the sensitive detection and identification of microorganisms in the THz regime is typically achieved by differentiating their dielectric properties, with the assistance of MMs. Yu et al. [89] functionalized $Fe_3O_4$@Au nanocomposites with *Staphylococcus aureus* specific aptamer for bacterial detection (Fig. 5a-b). Bacteria-nanomaterial complex could rapidly separate and enrich the target bacteria in a mixed bacterial solution, and the conjunction of nanomaterial significantly enhanced the resonance frequency shift when the samples are loaded on the THz MM sensor (Fig. 5c). Bacterial concentration monitoring and antibiotic susceptibility test can be achieved by a microfluidic MMs based on Fano resonance effect [90]. The detection limit for bacteria *E. coli* is as low as $5 \times 10^3$ cells/mL and bacterial concentration under antibiotic treatment could be dynamically obtained.

Recently, single bacterium sensing was achieved by THz scattering-type scanning near-field optical microscope (THz s-SNOM, Fig. 5d) [91]. Bacterial species demonstrate different morphological structures and THz near-field scattering intensities. The THz s-SNOM technique successfully distinguishes *E. coli* from *S. aureus* by the lower dielectric constant of *E. coli* compared to *S. aureus*, resulting in weaker scattered light (Fig. 5e-f). The observation can be well explained by the proposed finite dipole model. Antibiotic susceptibility test of *S. aureus* was further performed. Bacterial strains that are resistant to and susceptible to cell wall-inhibiting antibiotics produced distinct near-field signals. The classifying results obtained from THz-SNOM match well with traditional antibiotic susceptibility tests, while significantly reduces the time cost by its culture-free and label-free mechanism, presenting promising potential in drug-resistant bacteria screening.

The group from Ajou University proposes various approaches for the detection of living microorganisms based on THz waves. They fabricated SPR MMs to detect viable microorganisms, including molds, yeast and bacteria cells, with very low surface density (~0.05 cell/$\mu m^2$). Clear resonant frequency shifts were observed due to the change of the effective dielectric constant introduced by the microorganisms (Fig. 6a-b) [92], and the detection sensitivity can be further enhanced by employing the attenuated total reflection (ATR) geometry [93]. The chemical composition of the cell wall varies among different microorganisms. Therefore, they subsequently examined the main component in cell wall, and found that different peptidoglycan and polysaccharides result in different THz dielectric constants, leading to different MM response (Fig. 6c-d) [94]. Interestingly, in the subsequent work they further performed a thermal curve analysis with microorganisms-coated MMs and found unique THz fingerprint as a function of temperature [95, 96], rather than as a function of frequency. In detail, they heat up the MM sensors loaded with different bacteria and found the resonant frequency shows a sudden change at specific temperatures (Fig. 7a-c). These changes become more prominent in the first-derivative plot (Fig. 7d-e). Since the shifting temperatures are bacteria-specific, they could be leveraged as unique features to differentiate bacteria of different types. The characteristic temperatures were consistent with reported values measured by other approaches, which were classified into four temperature phases of growth phase, thermal inactivation phase, DNA denaturation phase and cell wall destruction phase. Traditional THz MM sensing only reflect the dielectric difference at the resonant frequency. This work took the thermal dimension into consideration. Pathogenic bacteria were easily distinguished from the bacterial mixture from the thermal curves. The work shows promising potential in bacterial identification in clinical and environmental settings.

3.2.3. Crop diseases detection

Crop diseases, mainly caused by fungal pathogens, significantly affect the yield and quality of agricultural products, which are highly related to food safety and human health. THz technology has been applied in the field of plant diseases diagnosis. Compared with time-consuming and expensive biochemical methods such as PCR and DNA microarray techniques, THz spectroscopy is advantageous in providing a wealth of information with low-cost and the capability of real-time in-site batch sampling. A typical example is the identification of fungal infections in chestnuts with 0.1 THz continuous-wave radiation [97]. The husk of chestnut is a thin and dry shell, and therefore transparent in the THz region, allowing non-destructive measurements without opening the outer shell. Inside the chestnut, the THz signals attenuate with the absorbance of the fruits based on Beer-Lambert's law (Fig. 8a-b). The total attenuation was found to be proportional to the water content, and infected chestnuts have reduced water content due to carbohydrates hydrolysis by fungi. Similar results were also found in hazelnuts [98]. Other plant diseases, including late blight and fusarium infection of potatoes [99], cucumber powdery mildew [66], tomato leaf mildew [100] and infested wood [101] were also investigated. In these studies, THz waves are able to assess the degree and depth of the diseased plant tissues and differentiate the decayed parts from the healthy parts (Fig. 8c-d). More recently, THz spectroscopy has been further combined with hyperspectral technology and machine vision [66, 100, 102] to enable a more comprehensive evaluation of plant health with better recognition accuracy.

3.2.4. Water pollution evaluation

THz spectroscopy has been further explored for water pollution evaluation. Microalgae are unicellular photosynthetic eukaryotes; they usually live in fresh water and oceans, producing approximately half of the oxygen on earth. Except for providing valuable metabolites like carotenoids, microalgae have emerging potential in bioremediation. For example, investigating the effects of heavy metal pollution (such as Lead ions $Pb^{2+}$ pollution) on microalgae provides important information for water quality management and algal bioremediation [103]. However, metabolites quantification of microalgae requires using several time-consuming techniques at the same time, such as chromatography and spectrophotometry. A simple, rapid and low-cost detection method is demanded. Because the main algal metabolites (including β-carotene, astaxanthin and starch) produced absorption signatures in the far to mid IR regime, the group led by Prof. PENG Yan has conducted a series studies of using FTIR spectrometer (0.9-20 THz) to quantify metabolites in algae in real time [104-106]. Generally, photosynthesis is inhibited under $Pb^{2+}$ stress and cells prefer to synthesis less-complex biomolecules (such as carbohydrate and carotenoid), rather than proteins. The measured THz characteristic frequency peaks correlate to the metabolite concentrations (Fig. 9). Their results showed that the storage of carbohydrate and carotenoid was facilitated. Based on the THz spectral data, the concentration of heavy metal ions can be well predicted by the established model with high accuracy, high efficiency and little sample amount. Similar technique was also applied in the heavy metal detection of soil [107]. These findings demonstrate the potential of using THz spectroscopy as a fast and non-destructive tool to evaluate the level of environmental pollution.

3.3. Viruses

Virus detection is always a critical topic for human health, which has attracted even more attentions after the coronavirus disease (COVID-19) pandemic. Numerous detection methods (e.g., real-time PCR) have been utilized for SARS-CoV-2 virus detection. However, faster and more accurate methods are continuously demanded, which can minimize the cost of treatment and isolation. THz spectroscopy has also been investigated as a tool for virus identification and characterization, especially for SARS-CoV-2. Most studies have used MMs [61, 62, 108-116] due to the significant scale mismatch between the virus/protein and the THz wavelength, as discussed in section 3.1.2. For example, Sengupta et al. [109] fabricated a MMs-based chip for coronavirus screening via exhaled breath analysis (Fig. 10a-b). This method lack robustness since all kinds of biological particles, including viruses, cytokines, cell debris and etc., can cause red-shifts of the resonance frequency of MMs. The authors claimed that shifts of about 1.5-9 GHz were observed for coronavirus positive patients, greater than shifts of about 0-1.5 GHz for healthy individuals.

Instead of sensing the virus directly, targeting the structural proteins of virus provides better specificity and robustness, which is more widely adopted. SARS-CoV-2 is formed by four major structural proteins, known as spike protein, envelope protein, membrane protein and nucleocapsid protein. Among them, spike protein has the largest size and locates at the outer surface of the virion. Spike protein is the first contact area to the host cell and responsible for viral entry (binding and fusing) into the host cell, hence it has been considered as an important diagnostic target. THz sensors have also been developed by targeting the spike protein for rapid and precise screening of SARS-CoV-2 [117]. The specific detection of spike protein was achieved by antibody [118], chemical treatment [119] and direct immersion in protein solution [120]. Derived peptides from the spike protein of different types of SARS viruses can be discriminated even within the same genus [121], showing great specificity. In addition, a low detection limit is demanded for efficient and massive COVID-19 screening. For this purpose, functionalized gold nanoparticles have been used to enhance the antibody binding strength and subsequent detection sensitivity (Fig. 10c-

d) [118]; magnetic nanoparticles were employed to selectively bind with the spike protein, and migrate together towards the THz MMs under external magnetic field [122]. The other structural proteins were less investigated [123]. Despite the promising accuracy demonstrated in different works, THz virus detections have only been conducted in laboratory settings. The robustness and anti-interference performance in in practical disease control applications still need to be validated through more comprehensive large-scale experiments.

3.4. Summary

This chapter reviewed interdisciplinary research linking THz technology with microbiological detection. THz studies on living microorganisms have been mainly focused on bacteria *E. coli* and *Bacillus* spores, fungi and microalgae. Other types of microorganisms, such as *Mycobacterium tuberculosis* [124, 125] and *Staphylococcus aureus* [64, 89, 126] and *P. aeruginosa* [78, 127, 128], protozoans of the genus *Trypanosoma* [129], were less investigated. For virus detection, coronavirus has received most attentions, while other infectious viruses have also been investigated either theoretically or experimentally, including Avian Infuenza (AI) viruses (e.g., H1N1, H5N2 and H9N2) [61, 62, 110-113, 130], bacteriophage [114, 115, 131], HIV virus [132] and Hepatitis B virus [116].

THz spectroscopy reveals certain degrees of dielectric difference between different microorganisms, although the contrast is moderate. It is usually necessary to amplify these differences using field-enhancing techniques, typically MMs. MM sensors have been broadly applied for prokaryotes, eukaryotes and viruses. This technique outperforms some commonly used biological methods by its relatively low cost, high sensitivity and rapid detection. However, the specificity against different microorganism species is still weak due to the lack of fingerprint absorptions in the THz range. THz thermal-spectroscopy extracts more unique features from the temperature domain. Although some works claimed that they have observed characteristic features, we highlight that these results are not supported by the condensed-matter theory and are more likely caused by errors or cavity oscillations. Therefore, developing binding techniques and combining them with THz sensors are important steps in the future to improve the target-specificity, promoting the application of THz microorganism detection.

## 4. Challenges and Outlooks

4.1. Intense water absorption

The strong absorption of water limits the available SNR and bandwidth in THz transmission measurements. To minimize the influence, dehydration process has been frequently applied. However, biomolecules and living matters are only viable under aqueous environment. It is important to use configurations more adaptive to hydrated biomaterials, such as defined-height sample chambers. The group from The Third Military Medical University of China did pioneering works in imaging living bacterial colony with such designs [78, 127]. Optical dielectric constants were measured for several common pathogenic bacterial species using a thickness-controllable sample chamber. They found that bacteria of different species or with different physiological states had varying hydration levels, and the small differences in water content led to distinguishable THz signals, which can be utilized for bacterial identification (Fig. 11a-b). Reflection configuration is another strategy to address the absorption issue [133]. In particular, attenuated total-internal reflection (ATR) provides the best sensitivity for absorbing materials [134]. Yu et al. applied the ATR technique for clinical samples from sputum, blood, urine and feces, and further used machine learning methods to analyze the spectrum (Fig. 11c-d). Bacterial stain isolated from different sample types and patient sources, or samples mixed with various biological components are highly heterogeneous. The results show that thirteen standard microorganisms could be rapidly recognized and accurately classified into three groups of gram-positive bacteria, gram-negative bacteria, and fungi. The total diagnostic accuracy reached 80.77%.

Microfluidic devices work in a way similar to the defined-height chamber to balance the water absorption and sensitivity [94, 135]. They offer another advantage of flexible input-output control that supports *in vitro* stimulus applications. Zhang et al. [136] developed a microfluidic cellular encapsulation device to measure the refractive index of living bacteria in near-physiological environments (Fig. 12). The viable cells were encapsulated in aqueous droplet at a moderate thickness. The physiological states of bacteria changed with external stimulus of copper ions $Cu^{2+}$, and they found the refractive index of bacterial droplet increased with increasing $Cu^{2+}$ concentration. The use of microfluidic device enables differentiating physiological states of bacteria under stress conditions. However, sensitive techniques for highly hydrated samples are still limited, especially for field-enhancing sensors. The resonant nature of most sensors suffer from the strong damping effect caused by the absorptive environment, significantly reducing their Q-factors. Reflection-type sensors with a shorter field-confining area may address these issues [137].

4.2. Poor spatial resolution

THz imaging suffers from poor spatial resolution due to the diffraction limit, especially at low frequencies. The far-field resolution is over ten times larger than a single cell, making them only applicable for bulk mixtures. The improvement factors of various far-field resolution-enhancing techniques are typically smaller than 10 [138-140], making them only applicable for large-scale samples. The detection of subtle changes inside microorganisms is nearly impossible, which hinders the sensitivity and specificity. Going beyond the diffraction limit requires near-field technologies. THz s-SNOM is an efficient approach to address this issue. The resolution of THz s-SNOM can reach down to a few tens of nanometers, which is fundamentally set by the tip radius rather than the wavelength of illumination. The technique utilizes a nanoscale tip oscillating at a certain frequency to scatter the THz near fields and demodulate it from the far-field signal based on the lock-in detection principle. The distance between the tip and the sample was precisely controlled based on the mechanism of atomic force microscope (AFM). In this way, both near-field THz dielectric information and three-dimensional morphology images can be obtained at the same time. THz s-SNOM has been used to investigate bacterial single cell and sub-cellular structures. Wang et al. [141] applied THz s-SNOM to measure single bacterial cells on a gold base. The contrast was significantly enhanced due to the large dielectric difference between the bacteria and the gold substrate. Clear morphology and fine inner structures were obtained in the THz images (Fig. 13a-c). The single cell of gram-negative bacteria *E. coli* and gram-positive bacteria *Staphylococcus aureus* can be differentiated. Zhang et al. [142] investigated cariogenic bacteria and related EPSs production using THz s-SNOM. Fine structures of cell and EPSs can be clearly obtained (Fig. 13d). The cost of this technique is the low SNR due to the weak near-field scattering, especially for biomedical samples which have very low dielectric permittivity.

4.3. Experimental standards and data reproducibility

Considering that THz biophotonics is a relatively new research area in the recent two to three decades, it is reasonable that standards have not been established for THz measurements. This is more difficult for biomedical applications because samples have varying types, forms and preparation methods. However, the lack of experimental standards makes it difficult to evaluate the data reliability and accuracy. Improper data processing results in errors, such as the spectral features claimed to have been observed in aqueous samples or mixtures of macromolecules. Another consequence is the low data reproducibility. Reported works are highly independent of each other, making data comparison nearly impossible. For example, studies investigating the radiation effects used different sources, power, exposure time, target samples and evaluation methods. Establishing standards in experimental setups, sample preparation and data processing, and encouraging more studies to reproduce and verify the existing works are important steps to improve the data quality in this field.

4.4. Outlooks

THz is a virgin frequency range with unique features and advantages in many interdisciplinary research areas. Microorganisms consist of a giant number of species that can adapt to a wide range of living environments, making them well-suited to the experimental conditions of THz measurements. Compared to animal tissues, the independent survival characteristics of microorganisms, along with their rich metabolic activity, are especially adaptive to the limited penetration depth of THz waves in hydrated samples. The interdisciplinary area between THz photonics and microorganisms has many interesting unexplored phenomenon, theories and applications to be revealed. Although many challenges are yet to be addressed, this research area attracts growing amounts of attention, with new findings and technologies being reported in recent years. With ongoing research, terahertz technology is expected to become a novel research tool or detection technique in the field of microbiology, promoting the advancement of both areas.

**Funding.** National Natural Science Foundation of China (62305069, 61988102); Natural Science Foundation of Guangdong Province (2024A1515010367); Talent Recruitment Project of Guangdong Province (2023QN10X375); Guangzhou Talent Recruitment Project (2024D01J0115).

**Disclosures.** The authors declare no competing interests.

**Tables**

Table 1. Experimental conditions for investigating the THz biological effects in microorganisms.

| Microorganism types | | THz source | THz type | Radiation frequency (THz) | Intensity (mW/cm²) | Exposure duration | Temperature (°C) | Biological effects | References |
|---|---|---|---|---|---|---|---|---|---|
| Bacteria | E. coli | Free electron laser | Pulsed, pulse duration 40-100 ps | 2.3 | 1400 | 15 min | / | Enhanced cell aggregation and cell adhesion; weakened cell division. | [31] |
| | E. coli | Free electron laser | Pulsed, pulse duration 2 ms | 3.1 | 33 | 8 h | 37 | Increased copy number of plasmid and protein production. | [32] |
| | E. coli | Free electron laser | Pulsed, pulse duration 5 ps | 4.0 ± 1.0 | / | 90 s | Room temperature | Significantly affects transcription process. | [33] |
| | Geobacillus icigianus | Free electron laser | Pulsed, pulse duration 50 ps | 2.3 | 230 | 15 min | 60 ± 1 | Various metabolic pathway affected (including cell growth, chemotaxis, etc.) | [35] |
| | E. coli | Free electron laser | Pulsed, pulse duration 50 ps | 1.5, 2.0 and 2.3 | 1400 | 15 min | 35 ± 2 | Enhanced protein expression (katG gene biosensor) | [48] |
| | E. coli | Free electron laser | Pulsed | 2.3 | 1400 | 15 min | 35 ± 2 | Enhanced protein expression (copA gene biosensor); no effect on emrR gene biosensor | [50] |
| Bacteria | E. coli | Free electron laser | Pulsed, pulse duration 100 ps | 2.3 | ~140 | 15-30 min | 36 ± 1 | | [51] |

| Microorganism types | | THz source | THz type | Radiation frequency (THz) | Intensity (mW/cm²) | Exposure duration | Temperature (℃) | Biological effects | References |
|---|---|---|---|---|---|---|---|---|---|
| Bacteria | E. coli | Solid-state device based on IMPATT-diode | Continuous wave | 0.14 | ~2.0 | | ~26 | Enhanced protein expression (*matA*, *safA* and *chbB* gene biosensor) | [52] |
| | | Free electron laser | Pulsed, pulse duration 100 ps | 2.3 | 140 (cuvette)/ 180 (microplate) | 15-30 min | 35-37 | Enhanced protein expression (*tdcR* gene biosensor) | |
| | | IMPATT-diode | Continuous wave | 0.14 | 2.0 | 15-30 min | 26 | | |
| | E. coli and Salmonella typhimurium | Free electron laser | Pulsed, pulse duration 50 ps | 2.3 | 1400 | 5-15 min | / | No mutagenicity and genotoxicity; positive effects on cell metabolism. | [37] |
| | Salmonella typhimurium and E. coli | Injection-seeded THz wave parametric generator | Pulsed | 1.6 | 3.8 | 20-60 min | 37 | No mutagenicity and DNA damage. | [38] |
| | E. coli | / | / | / | / | 15 min | / | No impact on viability and antimicrobial resistance | [54] |
| | Bacillus subtilis | Mechanically Tuned Gunn Oscillator | / | 0.094 | 1.3 | 1-24 h | 25 | No effect on metabolic activity or population density | [55] |

| Microorganism types | | THz source | THz type | Radiation frequency (THz) | Intensity (mW/cm²) | Exposure duration | Temperature (℃) | Biological effects | References |
|---|---|---|---|---|---|---|---|---|---|
| | E. coli | Two-stage gas THz laser | Pulsed, pulse duration 100 ns | 4.5 | / | 50-500 s | / | Cell death at a value of total energy of ~6 J. | [56] |
| Yeast | Saccharomyces cerevisiae | Backward wave oscillator | Continuous wave | 0.19-0.34 | ~5.78 | 30-150 min | 25 | Enhanced growth rate. | [39] |
| Archaea | Halorubrum saccharovorum | / | / | 2.3 | 800 | 5 h | / | Various protein expression level changed. | [40] |
| phytoplankton | Diatom algae | Free electron laser | Quasi-continuous, pulse duration 30-100 ps | 5.6 MHz (submillimeter wave) | 20000 | 3-10 s | / | Splitting of diatom frustules without destruction of cell content. | [41] |

Table 2. Parameters of MM sensors for microbial detection.

| Type | Analytes | Patten morphology | Deposition methods | Performance | Resonant frequencies | Note | References |
|---|---|---|---|---|---|---|---|
| Bacteria | E. coli | / | Capturing by phages | Limit of detection (LOD) $10^4$ CFU/mL | / | Based on suspended core THz fiber | [58] |
| | | / | Direct deposited | LOD $10^6$ CFU/mL | / | Based on metallic mesh sensor | [59] |
| | | Double-split ring resonator | Microfluidic device | LOD $5 \times 10^3$ CFU/mL, Q-factor 42 | 0.65 THz | Fano resonance effect | [90] |
| | E. coli S. aureus | Metal wire and a pair of split-ring resonators | Direct deposited | LOD ~$10^4$ CFU/mL; 378 GHz/RIU; Q-factor 21 | 1.53 THz | / | [126] |
| | S. aureus | Split-ring resonator | Specific aptamer binding | LOD $4.78 \times 10^2$ CFU/mL | 0.8 THz | Aptamer-functionalized Fe3O4@Au nanocomposites | [89] |
| | Cyanobacteria | Split-ring resonator | Direct deposited | / | 0.87 THz | Obtained a differential thermal curve | [96] |
| | Mycobacterium | / | / | Relative sensitivity of 90.6% | / | Based on photonic crystal fiber; only simulation | [124] |
| | | Bow-tie structure | Direct deposited | 1.5 THz/RIU; Q-factor 413 | 1.9, 2.7 THz | / | [125] |
| | Five bacterial strains | / | Direct deposited | / | / | Based on antenna | [60] |
| | Four bacterial strains | Split-ring resonator | Direct deposited | LOD 0.08 pg/mL | 0.86 THz | Based on gold nanoparticles and RCA | [64] |
| | Four bacterial strains | Hollow-core | Deposited on inner surface | / | 0.35, 0.5 THz | Based on photonic Bragg fiber | [128] |
| | Bacterial DNA | / | / | Genomic DNA LOD 0.05 ng/µL | / | Based on rolling circle amplification (RCA) | [63] |
| | | Asymmetry split-ring metasurface | Pyrene group binding | 100 nM DNA | 0.5 THz | Incorporated with microfluidic device | [135] |

| Type | Analytes | Patten morphology | Deposition methods | Performance | Resonant frequencies | Note | References |
|---|---|---|---|---|---|---|---|
| Molds, yeasts and bacteria | Molds, yeasts and bacteria | Split-ring resonator | Specific antibody binding | LOD $10^7$ units/mL | 0.84 THz | / | [92] |
| | 14 species from molds, yeasts and bacteria | Split-ring resonator | Microfluidic channel | / | 0.8 THz | / | [94] |
| | 10 species from yeasts and bacteria | Split-ring resonator | Direct deposited | 80 GHz/RIU | 0.77 THz | Obtained a differential thermal curve | [95] |
| | Yeast | Split-ring resonator | Direct deposited | LOD $7 \times 10^{-3}$ cell/$\mu m^2$; Q-factor 6 | 0.68 THz | ATR geometries | [93] |
| Viruses | Avian Influenza viruses | Jerusalem cross | / | / | 1.4, 3.2 THz | Only simulation; based on spoof surface plasmon polaritons | [61] |
| | | Grating split ring resonator | / | 300 GHz/RIU; Q-factor 690 | 1.93 THz | Based on THz surface plasmon polaritons | [62] |
| | | Nano-antenna | Direct deposited | / | 0.62, 0.93, 1.31 THz | / | [110] |
| | | H-shaped | / | 540 GHz/RIU | 1.72 THz | Only simulation; pattern material is graphene and substrate material is semiconductor | [111] |
| | | Chiral split ring | Specific antibody binding | ~4 dB/RIU | 1.15, 1.46 THz | Pattern material is graphene | [112] |
| | | Asymmetric split-ring resonators | Direct deposited | 30 GHz/RIU; Q-factor 6 | 0.4, 0.6 THz | / | [113] |
| | | Nanofake | Direct deposited | 9.2 GHz/RIU | 60 THz | Only simulation; pattern material is black phosphorus | [130] |

| Type | Analytes | Patten morphology | Deposition methods | Performance | Resonant frequencies | Note | References |
|---|---|---|---|---|---|---|---|
| Viruses | Flu viruses, SARS-CoV-2 virus | Star-shaped holes | Direct deposited | 2200 GHz/RIU; Q-factor 19 | 1.97, 3.37 THz | / | [108] |
| | SARS-CoV-2 virus | Cross-arrowhead | Breath exhaled | / | 0.81 THz | / | [109] |
| | | Split-ring resonator | / | 490 GHz/RIU | 2.3 THz | Only simulation; pattern material is graphene | [117] |
| | SARS-CoV-2 virus spike protein | Toroidal metasurface | Specific antibody binding | LOD ~4.2 fM; Q-factor 14 | 0.4, 0.6 THz | AuNPs functionalized | [118] |
| | | Three-split ring | Direct deposited | LOD 5 ng; 73.2 GHz/RIU | 0.68, 1.63 THz | / | [119] |
| | | Split-ring resonator | Immersion | / | 0.85, 1.06 THz | / | [120] |
| | SARS-CoV-2 virus spike protein | Elliptical grooves | Specific antibody binding | LOD 0.002 ng/mL | 0.53 THz | Utilized magnetic nanoparticles | [122] |
| | SARS-CoV-2 spike protein-derived peptides | Nanoslot arrays | Direct deposited | LOD 0.1 mg/mL (i.e., 41.7 µM). | 1.16/1.64/2.07 THz | / | [121] |
| | Bacteriophage | Split-ring resonators | Direct deposited | 70 GHz/RIU | 0.8/1.2 THz | 200 nm gap | [114] |
| | | Nanogap-loop array | Direct deposited | / | 0.77 THz | Virus-sized nanogap | [115] |
| | | Hybrid slot antenna | Spin-coated | 32.7 GHz·µm$^2$/particle | 0.7 THz | Pattern material is gold layer with silver nanowires | [131] |

| Type | Analytes | Patten morphology | Deposition methods | Performance | Resonant frequencies | Note | References |
|---|---|---|---|---|---|---|---|
| Viruses | Hepatitis B virus DNA | Split-ring resonator | Direct deposited | LOD 127 IU/mL | 0.95 THz | / | [116] |
| | Viruses HSV, HIV-I, and M13 | L-shaped | Direct deposited | 1012 GHz/RIU | 4.5 THz | Pattern material is InAs; polyamide film in the middle and gold substrate at bottom | [132] |
| Eukaryotes | *Trypanosomes* | Asymmetric double-split ring resonator | Specific aptamer binding | / | / | / | [129] |

**Figures**

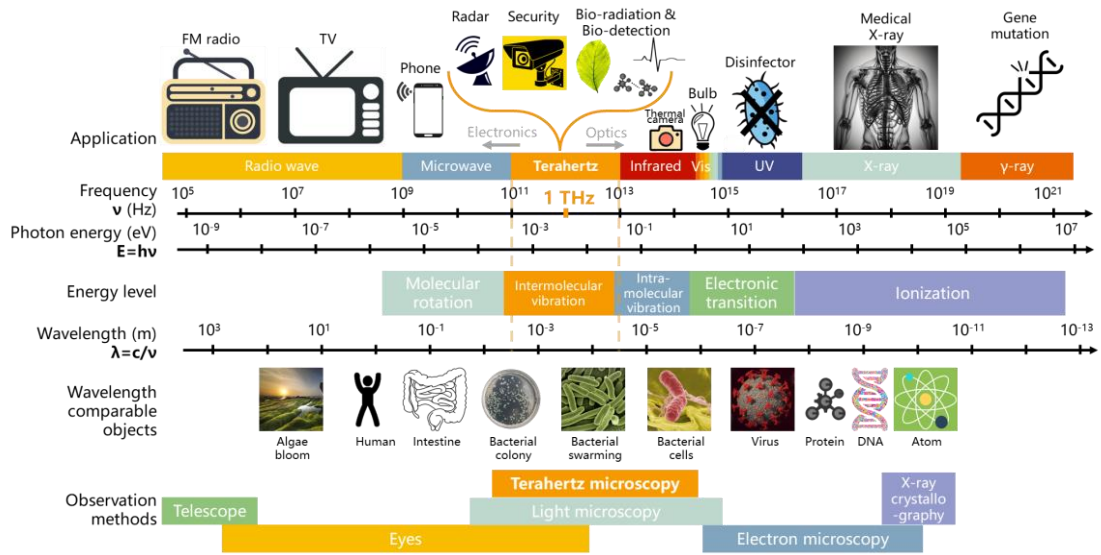

Figure. 1. Comparison of electromagnetic wave spectrum and size of microorganism and their sub-cellular components. Images are found on *pixabay.com* with permission.

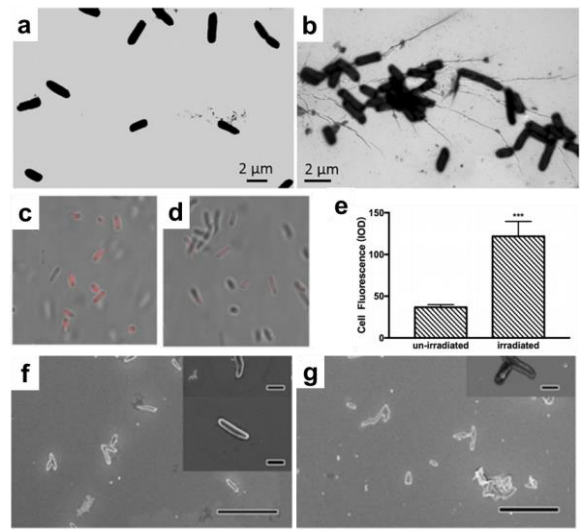

Figure. 2. (a) Stand-alone bacteria before irradiation and (b) aggregated bacteria after the irradiation. Panels a,b are reproduced with permission from Peltek *et al.*, Sci. Rep. **11**, 20464 (2021) [31]. Copyright 2021 Authors, licensed under a Creative Commons Attribution (CC BY) license. (c) Merged optical images and fluorescent images of *E. coli* cells after exposure to THz radiation and (d) the untreated control. (e) Quantification of fluorescence intensity of *E. coli* cells after exposure to THz radiation. Panels c-e are reproduced with permission from Zhao *et al.*, Biomed. Opt. Express **11**, 3890 (2020) [32]. Copyright 2020 Optical Society of America. (f) Following 10 min of THz exposure from a synchrotron source, *E. coli* cells exhibited a dehydrated appearance, accompanied by cytosolic leakages. (g) At 90 min of THz exposure, *E. coli* cells had altered morphology. Panels f,g are reproduced with permission from Ivanova *et al.*, ACS omega **9**, 49878 (2024) [34]. Copyright 2024 Authors, licensed under a Creative Commons Attribution (CC BY) license.

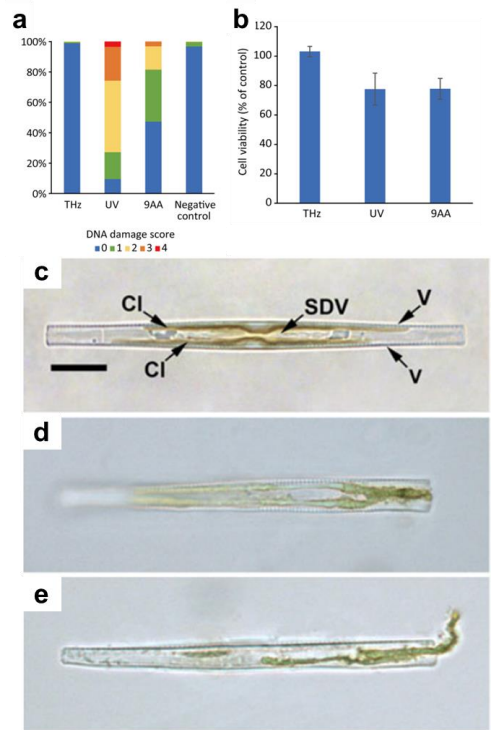

Figure. 3. (a) Proportion of cells with different DNA damage scores. THz laser irradiation group and the negative control group were mostly composed of cells with score 0 (undamaged), whereas 90% of UV-irradiated cells and 53% of 9AA-treated cells were scored 1 or higher, indicating that the DNA was damaged. (b) Cell viability (% of control) after treatment with THz, UV, and 9AA. The values and error bars indicate the means and standard deviations, respectively (n=3 for each group). THz laser irradiation did not affect cell viability, whereas the treatments with UV and 9AA reduced cell viability to approximately 80%. Panels a,b are reproduced with permission from Shirato *et al.*, Photochem. Photobiol. **100**, 146 (2023) [38]. Copyright 2023 John Wiley and Sons. Light microscopy images of diatom cells (c) before and (d,e) after action of high-power sources of submillimeter laser irradiation. The labels in (c) are as follows: Cl chloroplasts, V valves of the parent cell, SDV two silica deposition vesicles in which new valves are formed. Panels c-e are reproduced with permission from Annenkov *et al.*, Eur. Biophys. J. **42**, 587 (2013) [41]. Copyright 2013 Springer Nature.

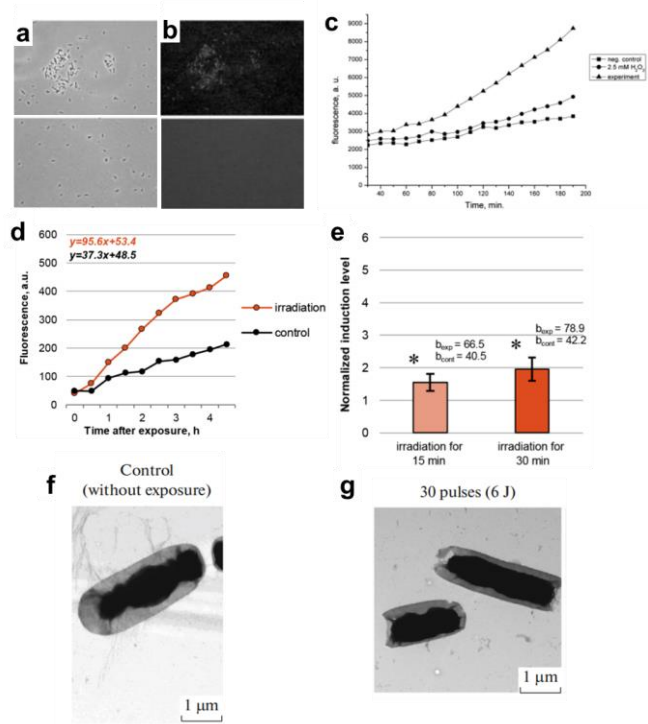

Figure. 4. *E. coli/pKatG-gfp* biosensor cells after exposure to THz radiation at a wavelength of 130 μm for 15 min. (a) Phase contrast picture of irradiated cells (up panel) and control cells (bottom panel); (b) fluorescence microscopy picture of the irradiated cells (up panel) and control cells (bottom panel); (c) Fluorescence intensity of the GFP in *E.coli/pKatG-gfp* biosensor cells after 15 min exposure to THz radiation at a wavelength of 130 μm, normalized to the background fluorescence values. Panels a-c are reproduced with permission from Demidova *et al.*, Bioelectromagnetics **34**, 15 (2013) [48]. Copyright 2012 John Wiley and Sons. (d) Typical dynamics of the fluorescence in response to 30 min THz irradiation in comparison with a control (results of one independent replicate are presented — fluorescence curves and the respective linear regression equations); (e) normalized induction levels (average values from six biological replicates) at 4.5 h after the exposure. The error bars represent standard deviation; $b_{exp}$ and $b_{cont}$ are the average slope coefficients in experiment and control, respectively. *Significant differences ($P < 0.05$) in slope coefficients between experiment and control. Panels d,e are reproduced with permission from Serdyukov *et al.*, Biomed. Opt. Express **12**, 705 (2021) [52]. Copyright 2021 Optical Society of America. (f) Electron microphotographs of the *E. coli* cells without being exposed to THz irradiation. (g) Exposure to approximately 6 J of total THz wave energy could damage bacterial cell and cause cell death. Panels f,g are reproduced with permission from Boev *et al.*, Biophysics **64**, 416 (2019) [56]. Copyright 2019 Springer Nature.

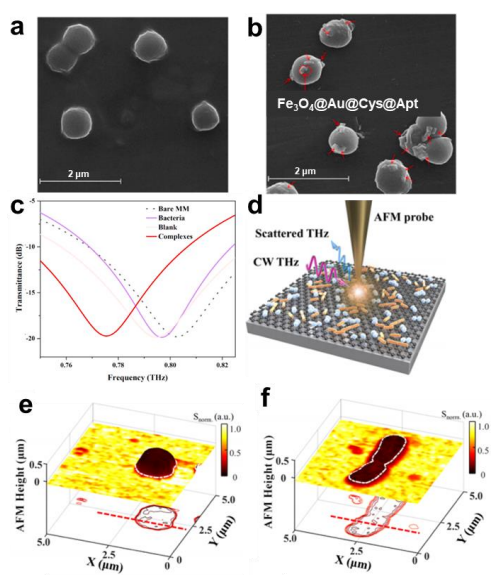

Figure. 5. TEM images of *S. aureus* (a) before and (b) after captured by Fe$_3$O$_4$@Au@Cys@Apt. (c) Normalized THz transmission spectra of different samples: Bare MMs: bare metamaterials; Bacteria: *S. aureus* standard solution at $1 \times 10^8$ CFU/mL; Blank: Fe$_3$O$_4$@Au@Cys@Apt plus ultrapure water without bacteria; Complexes: nanoparticle-bacteria complexes. Panels a-c are reproduced with permission from Yu *et al.*, Talanta **272**, 125760 (2024) [89]. Copyright 2024 Elsevier. (d) Schematic of the customized THz s-SNOM platform for single-bacterium THz nanoimaging. (e-f) near-field intensity S$_{norm}$ color-coded onto the topography of single (e) *S. aureus* and (f) *E. coli* (recorded at 0.1 THz). The corresponding contour plots represent bacterial profiles with near-field intensities ranging from 0 to 0.03 a.u., and the contour interval is 0.006 a.u. Panels d-f are reproduced with permission from Zhou *et al.*, ACS Appl. Mater. Interfaces **17**, 18074 (2025) [91]. Copyright 2025 American Chemical Society.

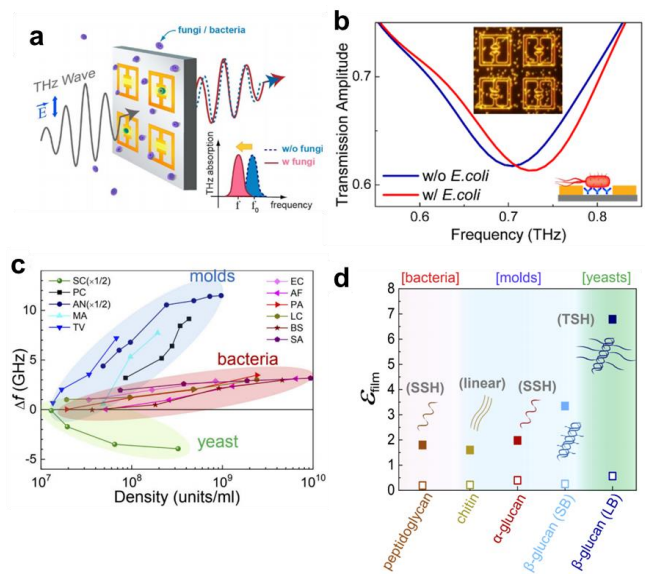

Figure. 6. (a) Schematic presentation of THz metamaterials sensing of microorganisms. (b) THz transmission before (blue line) and after (red line) the deposition of *E. coli* on the functionalized metamaterials in aqueous environments. (inset) Corresponding dark-field microscopic image obtained after the deposition of *E. coli*. Panels a,b are reproduced with permission from Park *et al.*, Sci. Rep. **4**, 4988 (2014) [92]. Copyright 2014 Springer Nature. (c) Plot of frequency shift as a function of number density for different species. (d) Real (filled boxes) and imaginary (open boxes) parts of dielectric constants for peptidoglycan and polysaccharides films, measured at 1 THz. Panels c,d are reproduced with permission from Yoon *et al.*, Biomed. Opt. Express **11**, 406 (2019) [94]. Copyright 2019 Optical Society of America.

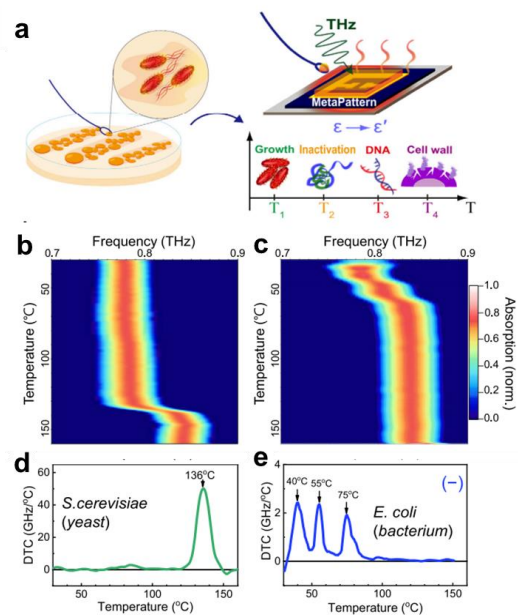

Figure. 7. (a) Schematic of the THz experiments. The microbial films grown on a culture medium are transferred to THz metamaterials on a ceramic heater. The ceramic heater is punctured at the center with a diameter of 2 mm to enable the transmission experiments. The microbes exhibit phase change with increasing temperature at conditions of growth, inactivation, DNA denaturation, and cell wall destruction. (b,c) 2D plot of THz absorption through metamaterials coated with a yeast layer (*S. cerevisiae*) and *E. coli* layer as functions of THz frequency (x-axis) and temperature (y-axis). (d,e) Differential thermal curves for yeast and *E. coli* obtained from panel (b) and (c) . Minus sign (-) in (e) indicates that *E. coli* is Gram-negative bacteria. Panels a-e are reproduced with permission from Jun *et al.*, Nat. Commun. **13**, 3470 (2022) [95]. Copyright 2022 Authors, licensed under a Creative Commons Attribution (CC BY) license.

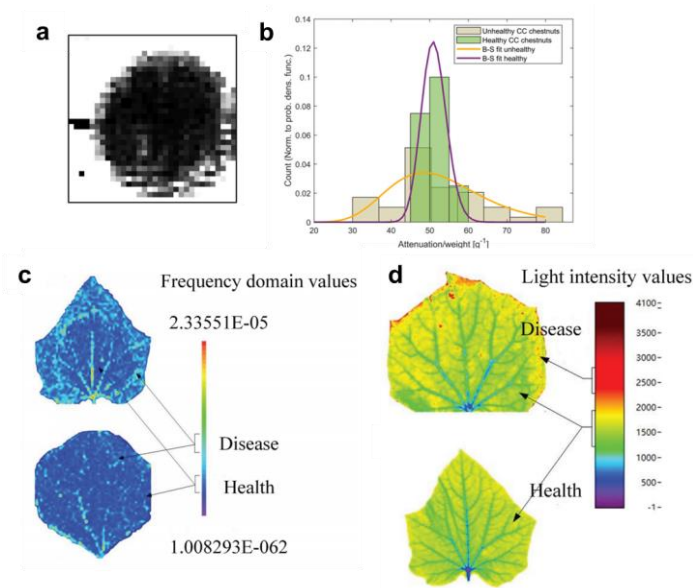

Figure. 8. (a) An example of a black and white image of the chestnut obtained by the original data matrix. (b) Histograms of overall attenuation/weight of healthy and unhealthy chestnuts, fitted by a Birnbaum-Saunders distribution. Panels a,b are reproduced with permission from Di Girolamo *et al.*, Food Control **123**, 107700 (2021) [97]. Copyright 2021 Elsevier. (c) THz characteristic images and (d) NIR characteristic images at 1395 nm of cucumber powdery mildew. Panels c,d are reproduced with permission from Zhang *et al.*, Front. Plant Sci. **13**, 1035731 (2022) [66]. Copyright 2022 Authors, licensed under a Creative Commons Attribution (CC BY) license.

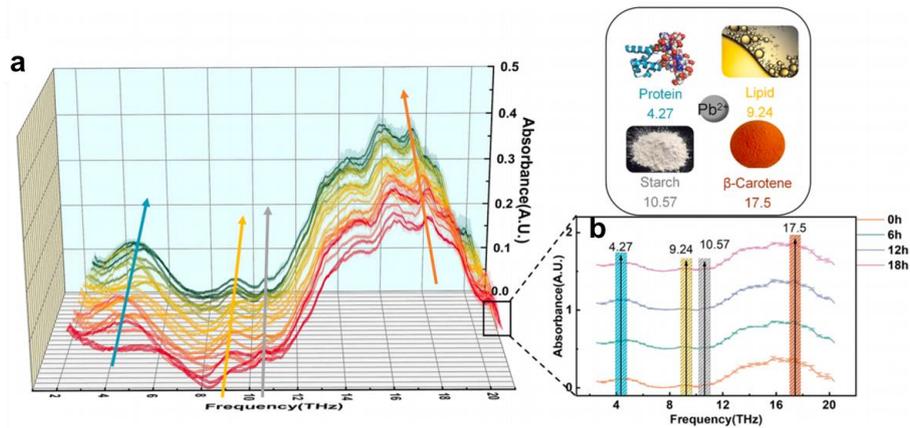

Figure. 9. (a) shows the spectra of microalgae under the stress of seven different $Pb^{2+}$ concentrations at 0 h, 6 h, 12 h and 18 h. (b) shows the spectra of microalgae under the stress of 5 × $10^3$ ng/mL $Pb^{2+}$ concentrations at 0 h, 6 h, 12 h and 18 h. Panels a,b are reproduced with permission from Shao *et al.*, J. Hazard. Mater. **435**, 129028 (2022) [106]. Copyright 2022 Elsevier.

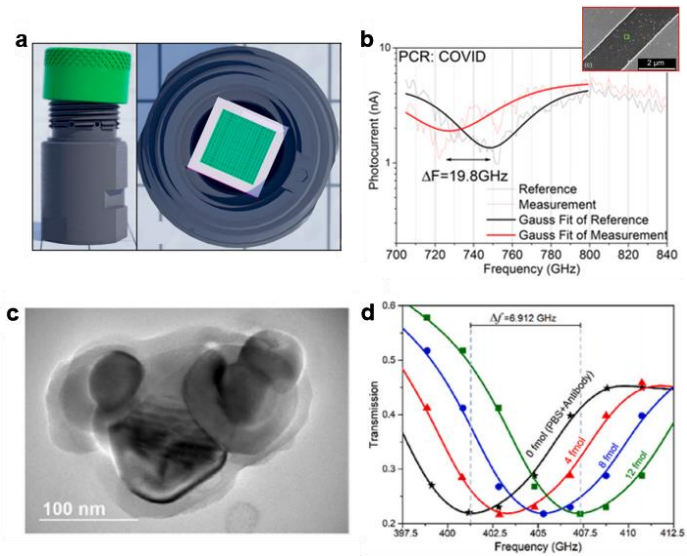

Figure. 10. (a) Conceptual design of the COVID-19 breathalyzer test. The capsule can be placed in a THz spectrometer for testing after external sterilization. (b) Transmittance spectra showing ΔF for an Au metamaterial chip collected from a COVID-19-infected patient. Insert: SEM images of the metamaterial chip. Panels a,b are reproduced with permission from Sengupta, *et al.* ACS Appl. Nano Mater. **5,** 5803 (2022) [109]. Copyright 2022 American Chemical Society. (c) A TEM image of AuNPs conjugated with the respective antibody of SARS-CoV-2. (d) Measured transmission spectra of the THz meta sensor device for different concentrations (4–12 fM) of SARS-CoV-2 spike protein. For 0 fM of spike protein, the spectral response was defined merely for antibody conjugated AuNPs plus PBS. Panels c,d are reproduced with permission from Ahmadivand *et al.*, Biosens. Bioelectron. **177**, 112971 (2021) [118]. Copyright 2021 Elsevier.

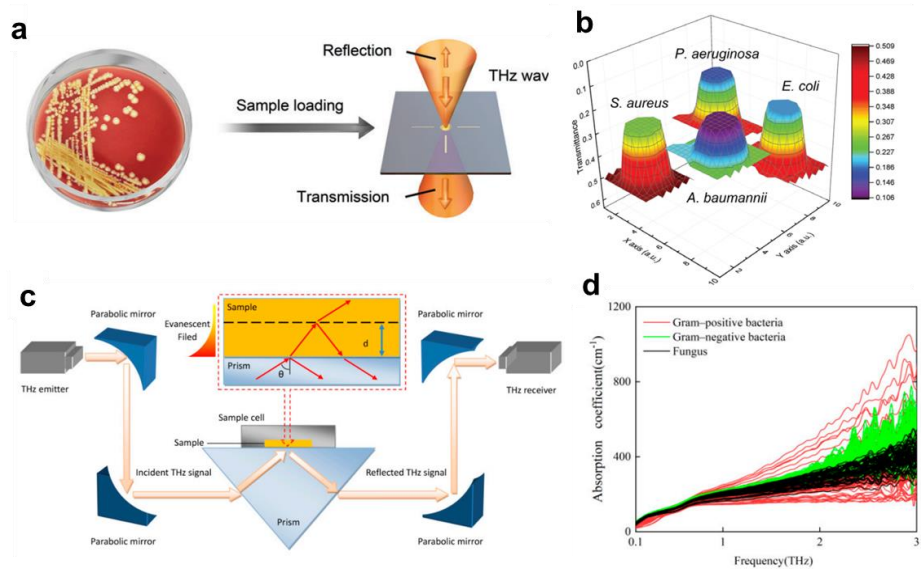

Figure. 11. (a) Schematic of sample loading and THz measurement of different bacterial colonies. (b) Comparison of THz images for different bacteria in stereo view. Note that 4 independent tests for different bacteria were drawn in the same coordinate system for easy comparison. Panels a,b are reproduced with permission from Yang *et al.*, J. Biophotonics **11**, e201700386 (2018) [127]. Copyright 2018 John Wiley and Sons. Identification of clinical microbes based on THz-ATR spectra. (c) Schematic illustration of the THz-ATR spectrometer with a sample cell made of Si. Inset: Diagram of the "prism–sample" model. (d) THz absorption spectra of eight Gram-positive bacterial strains (red), two Gram-negative bacterial strains (green), and three fungi (black). Panels c,d are reproduced with permission from Yu *et al.*, Biosensors **12**, 378 (2022) [134]. Copyright 2022 Authors, licensed under a Creative Commons Attribution (CC BY) license.

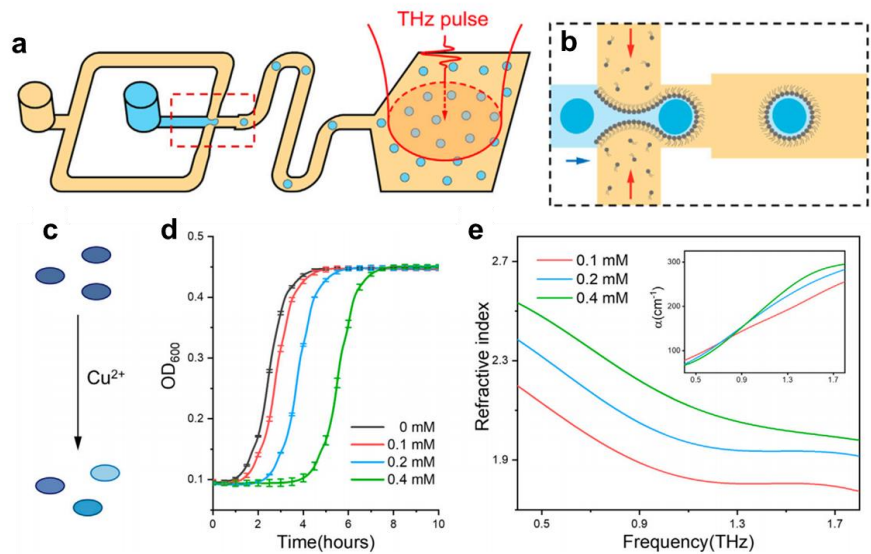

Figure. 12. The integration of droplet microfluidics for automated and high-through performance. (a) Schematic of the microfluidic device that can generate droplet sample for THz-TDS measurement. An aqueous phase containing cells passes through a flow-focusing junction where it meets an oil phase containing lipids, leading to the generation of droplets. Red arrows show the position of a cell before and after encapsulation. Droplets are collected in a quartz chamber which serves as the THz measurement cuvette. (b) The schematic of the droplet generation process. (c) Schematic showing $Cu^{2+}$ ions treatment induces distinct activity states of bacterial cells. (d) The growth curve of E. coli under 0, .1, .2 and .4 mM of $Cu^{2+}$ treatment, respectively. (e) Refractive index and absorption coefficient spectra (inset) of *E. coli* cells. All bacterial samples were treated with $Cu^{2+}$ and accounted to the same cell numbers for THz measurement. Panels a-e are reproduced with permission from Zhang *et al.*, Front. bioeng. biotechnol. **10**, 1105249 (2023) [136]. Copyright 2023 Authors, licensed under a Creative Commons Attribution (CC BY) license.

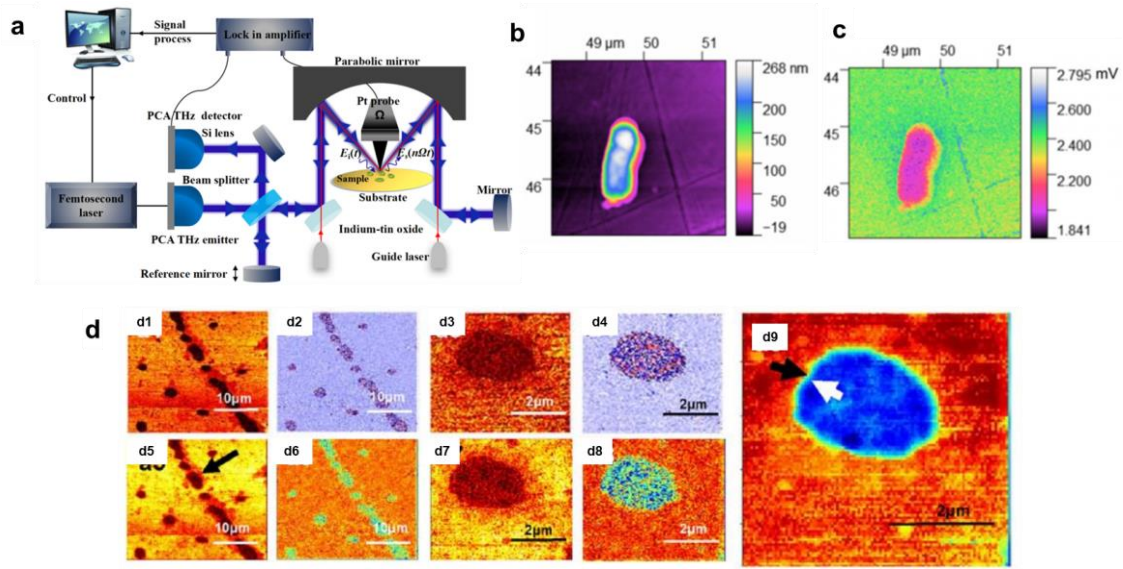

Figure. 13. (a) Schematic illustration of the THz s-SNOM setup and its use for single bacteria imaging. (b) AFM topographic map of *E. coli*. (c) THz near-field amplitude image of *E. coli*. Panels a-c are reproduced with permission from Wang *et al.*, Front. Microbiol. **14**, 1195448 (2023) [141]. Copyright 2023 Authors, licensed under a Creative Commons Attribution (CC BY) license. (d) Processed THz s-SNOM images of *S. mutans,*. where (d1-d4) are before and (d5-d8) and after processing, respectively; d9 is the image after repeated iterative processing of d8; black arrows represent the extracellular matrix; white arrows represent the cell wall. Panel d is reproduced with permission from Zhang *et al.*, J. Dent. Res. **103**, 1428 (2024) [142]. Copyright 2024 Sage Publications.